\documentclass[aps,pra,twocolumn,reprint,superscriptaddress,nofootinbib,amsmath,amssymb,amsfonts]{revtex4-2}
\usepackage{amssymb}
\usepackage{amsmath}
\usepackage{amsthm}
\usepackage{bm}
\usepackage{graphicx}
\usepackage{float}
\usepackage[hyperfootnotes=false]{hyperref}
\usepackage{tabularx}
\usepackage{dcolumn}
\usepackage{isotope}
\usepackage{xcolor}
\usepackage{bbold}
\usepackage{lipsum}

\usepackage{afterpage}
\usepackage{multirow}
\providecommand{\ket}[1]{|#1\rangle}

\newcommand{\RED}[1]{\textcolor{red}{#1}}
\newcommand{\BLUE}[1]{\textcolor{blue}{#1}}
\newcommand{\GREEN}[1]{\textcolor{teal}{#1}}
\newcommand{\VIOLET}[1]{\textcolor{violet}{#1}}

\newcommand{\GRAY}[1]{\textcolor{darkgray}{#1}}
\newcommand{\MAG}[1]{\textcolor{magenta}{#1}}
\newcommand{\ds}{\boxplus}
\newcommand{\rd}{\oplus}

\usepackage{bbm}
\usepackage{dsfont}
\newcommand{\unity}{\mathbbmss{1}}
\newcommand{\target}{\mathcal{H}}
\newcommand{\gens}{\mathcal{G}}
\newcommand{\lie}[1]{\langle #1 \rangle_{\text{Lie}}}
\newcommand{\g}{\mathfrak{g}}
\newcommand{\Z}{\mathcal{Z}}
\newcommand{\h}{\mathfrak{h}}
\newcommand{\C}{\mathds{C}}
\newcommand{\R}{\mathds{R}}
\newcommand{\su}{\mathfrak{su}}
\newcommand{\uu}{\mathfrak{u}}
\newcommand{\so}{\mathfrak{so}}
\newcommand{\usp}{\mathfrak{sp}}
\newcommand{\SU}{\mathrm{SU}}

\usepackage{physics}

\theoremstyle{definition}

\begin{document}

\title{\texorpdfstring{Ground-state reachability for variational quantum eigensolvers:\\ a Rydberg-atom case study}{Ground-state reachability for variational quantum eigensolvers: a Rydberg-atom case study}}

\author{Juhi~Singh\,\href{https://orcid.org/0000-0001-9807-9551}{\includegraphics[height=6pt]{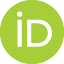}}}
\email[]{j.singh@fz-juelich.de}
\affiliation{Forschungszentrum Jülich GmbH, Peter Grünberg Institute, Quantum Control (PGI-8), 52425 Jülich, Germany}
\affiliation{Institute for Theoretical Physics, University of Cologne, 50937 Köln, Germany}

\author{Andreas Kruckenhauser\,\href{https://orcid.org/0000-0002-2420-648X}{\includegraphics[height=6pt]{ORCID-iD_icon-64x64.png}}}
\email[]{andreas.kruckenhauser@planqc.eu}
\affiliation{Institute for Theoretical Physics, University of Innsbruck, 6020 Innsbruck, Austria}
\affiliation{Institute for Quantum Optics and Quantum Information of the Austrian Academy of Sciences, 6020 Innsbruck, Austria}
\affiliation{PlanQC GmbH, 85748 Garching, Germany}

\author{Rick van Bijnen\,\href{https://orcid.org/0000-0002-0979-2521}{\includegraphics[height=6pt]{ORCID-iD_icon-64x64.png}}}
\email[]{rick@planqc.eu}
\affiliation{Institute for Theoretical Physics, University of Innsbruck, 6020 Innsbruck, Austria}
\affiliation{Institute for Quantum Optics and Quantum Information of the Austrian Academy of Sciences, 6020 Innsbruck, Austria}
\affiliation{PlanQC GmbH, 85748 Garching, Germany}

\author{Robert~Zeier\,\href{https://orcid.org/0000-0002-2929-612X}{\includegraphics[height=6pt]{ORCID-iD_icon-64x64.png}}}
\email[]{r.zeier@fz-juelich.de}
\affiliation{Forschungszentrum Jülich GmbH, Peter Grünberg Institute, Quantum Control (PGI-8), 52425 Jülich, Germany}

\date{June 27, 2025}

\begin{abstract}
As quantum computing progresses, variational quantum eigensolvers (VQE) 
for ground-state preparation have become an attractive option in
leveraging current quantum hardware. However, a major challenge in implementing VQE
is understanding whether a given quantum system can even reach the target ground state.
In this work, we study reachability conditions for VQE by  
analyzing their inherent symmetries.
We consider
a Rydberg-atom quantum simulator with global controls and evaluate its ability to reach ground states for Ising and Heisenberg target Hamiltonians.
Symmetry-based conclusions for a smaller number of qubits are corroborated by VQE simulations, demonstrating the reliability
of our approach in predicting whether a given quantum architecture could successfully reach the ground state. 
Our framework also suggests approaches to overcome symmetry restrictions
by adding additional quantum resources or choosing different initial states, offering practical guidance for implementing
VQE in quantum simulation architectures. Finally, we illustrate connections to adiabatic state preparation.
\end{abstract}

\maketitle

\section{Introduction\label{sec:introduction}}
One promising solution to address limitations of classical computing is to apply quantum co-processors, which
process information at the quantum level using qubits instead of digital bits
\cite{feynman_simulating_1982, abrams_simulation_1997,arute_quantum_2019}. This enables us to
leverage superposition and entanglement to eventually outperform classical
algorithms \cite{noauthor_polynomial-time_nodate}. 
Although ideal quantum processors are still developing, imperfect devices from fifty to thousand qubits have
demonstrated intriguing performance for certain tasks, marking the start of the noisy intermediate scale quantum
(NISQ) era \cite{preskill_quantum_2018, arute_quantum_2019, bharti_noisy_2022,mcardle_quantum_2020, zhong_qca_2020, zhu_zuchongzhi_2021, madsen_photonic_2022, kim_utility_2023, Miessen_2024, liu_certified_2025, king_beyondclassical_2025}. 

On the software front, various research groups have been developing effective algorithms applicable to quantum computing and simulation.
Among these are variational quantum algorithms, which enable the use of current quantum hardware by integrating with classical
optimization techniques \cite{peruzzo_variational_2014, farhi_2014_QAOA, cerezo_variational_2021,endo_hybrid_2021,cao_quantum_2019}. These algorithms leverage a
parameterized quantum circuit to prepare trial wavefunctions, which are iteratively adjusted in a feedback loop with a classical
computer minimizing a cost function reflecting the quality of a prepared quantum state.
Variational quantum eigensolvers (VQE)
\cite{peruzzo_variational_2014, mcclean_theory_2016, wecker_progress_2015,tilly_variational_2022} aim at preparing a good
approximation of the ground state of an Hamiltonian, while using the energy of the state as a cost function. VQE 
optimizes variational parameters of a quantum circuit 
based on a sequence of quantum logic gates
for digital quantum simulators \cite{moll_quantum_2018, hempel_quantum_2018}.
Different ansätze have been proposed and analyzed for variational optimization tasks
\cite{mcclean_barren_2018, cerezo_challenges_2022,  cerezo_cost_2021, marciniak_optimal_2022, kazi_analyzing_2024}. 
Moreover, for analog quantum simulators, multiple unitary blocks are driven using the parameterized controls
\cite{kandala_hardware-efficient_2017, kokail_2019, darcangelo_leveraging_2024} which are variationally adjusted.
These algorithms are particularly intriguing within the framework of quantum phase transitions \cite{kokail_2019, Meth_2022, kattemolle_variational_2022, okada_classically_2023}. After reaching the ground state,
one can experimentally analyze its characteristics, such as correlation functions, and investigate how the system responds to externally applied perturbations.

Variational algorithms have been successfully implemented on various platforms, including photonic processors \cite{peruzzo_variational_2014, cimini_variational_2024}, superconducting qubits
\cite{google_ai_quantum_and_collaborators_hartree-fock_2020, kandala_error_2019, kandala_hardware-efficient_2017}, and trapped ion systems \cite{kokail_2019, pagano_2020, nam_ground-state_2020, marciniak_optimal_2022, Meth_2022}.
Alongside these platforms, neutral atoms excited to Rydberg states gained attention as a promising quantum processing platform, offering coherent control \cite{bluvstein_logical_2024, evered_high-fidelity_2023, levine_parallel_2019, peper2025spectroscopy, senoo2025high} over hundreds of  atoms \cite{scholl2021quantum, ebadi2021quantum, manetsch2024tweezer, tao2024high, gyger_continuous_2024, park2022cavity} trapped in reconfigurable optical arrays \cite{barredo2018synthetic, semeghini2021probing, bluvstein2022quantum}.
In particular, these systems can be controlled via global and/or local external fields, enabling the execution of parameterized quantum circuits suitable for VQE applications \cite{michel_blueprint_2023, shaw_benchmarking_2024}.
However, it is generally difficult to predict whether a given set of control parameters is sufficient to prepare the ground state of a particular target Hamiltonian \cite{larocca_review_2024}.

Control theory has been extensively applied both experimentally and theoretically to address such challenges
\cite{keijzer_pulse_2023, PRXQuantum.2.010101, kurniawan_controllability_2012}. By using Lie-algebraic principles, control
theory offers a robust framework for characterizing operations that a quantum device can perform
\cite{jurdjevic_control_1972, dalessandro2022,  zimboras_dynamic_2014}. The Lie algebra is used to study controllability, simulability,
and reachability conditions for different quantum devices \cite{ramakrishna_controllability_1995,schirmer_complete_2001,altafini_controllability_2002,schirmer_identification_2002,polack_uncontrollable_2009,schirmer_degrees_2002,albertini_controllability_2018,schirmer_criteria_2002,zeier_symmetry_2011,zimboras_dynamic_2014,zeier2015squares,zimboras_symmetry_2015,schulte2017,schulte2018,kazi_analyzing_2024}.
However, calculating the Lie algebra becomes difficult
for increasing system sizes. For larger number of qubits, we can still determine and study the symmetries of a quantum system, i.e., 
the matrices that commute with the considered Hamiltonians \cite{zeier_symmetry_2011,zimboras_symmetry_2015}.

In this work, we apply and extend
the Lie-algebra and symmetry tools developed in
\cite{zeier_symmetry_2011,zimboras_dynamic_2014,zeier2015squares,zimboras_symmetry_2015,schulte2017,schulte2018, kokail_2019, kazi_analyzing_2024}
to match available quantum resources with the desired target ground states, assessing whether a given variational circuit can,
in principle, prepare the ground state of a given target Hamiltonian. 
We focus on the ground state preparation for an Rydberg-atom analog simulator with global controls.
We first consider the Lie-algebra and simulability structure of the resource Hamiltonians without and with the target Hamiltonian,
which reveals that the ability to simulate a target Hamiltonian does not imply that one can reach its ground state(s).
For a complete analysis, we determine the invariant subspaces of the resource Hamiltonians highlighting that the success of VQE
depends critically on the position of the initial state and the target ground state in the invariant subspaces. 
We relate our symmetry analysis for testing the ground state reachability
to adiabatic state preparation \cite{farhi2000} and the adiabatic theorem \cite{BornFock1928,Kato1950,Messiah1962,teufel2003,Teufel2022,Simon2019}.
Lastly, we illustrate how symmetries can guide us in preparing a suitable initial state to reach the ground state of the target Hamiltonian. 
This study provides tools aimed at analyzing and improving the execution of
variational quantum algorithms.

The work is structured as follows:
In Sec.~\ref{sec:VQE}, we describe the considered Rydberg-atom quantum simulator and its resource Hamiltonians
and we detail the VQE approach based on the example of Ising and Heisenberg target Hamiltonians. 
In Section~\ref{sec:tools}, we introduce tools for the analysis of the reachability for VQE. In particular,
Sec.~\ref{symmetry_analysis} contains our symmetry analysis based on invariant subspaces.
Section~\ref{sec:larger_system} extends our symmetry analysis to up to
ten qubits. Different initial states and their VQE performance are considered
in Sec.~\ref{sec:initial_state}.  Section~\ref{sec:adiabatic} discusses connections to adiabatic state preparation.
Finally, we present our conclusions in Sec.~\ref{sec:conclusion}. Certain aspects are deferred to appendices.

\section{Variational quantum eigensolver (VQE)\label{sec:VQE}}
\subsection{Resource Hamiltonians \label{sec:VQE:model}}

Neutral atoms excited to Rydberg states can act as an analog quantum simulator, capable of executing parameterized quantum circuits.
In particular, we consider a setup where external control fields act globally on all atoms, thus a parametrized quantum circuit can be implemented by modulating control field parameters over time.
Furthermore, the setup consists of $N$ neutral atoms trapped in an optical lattice or tweezer array, with the qubit being encoded in a ground and Rydberg state of each atom \cite{robicheaux2005many, weimer2008quantum, bernien2017probing, choi2023preparing}.
For such an encoding the interactions between a pair of atoms in the Rydberg state are typically of van der Waals type (long-range) and always present.

\begin{figure}
\includegraphics{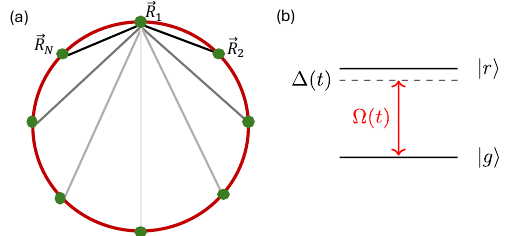}
\caption{(a) Circular geometry: Illustration of the spatial positioning of laser trapped neutral atoms (green dots). Atoms excited to Rydberg states experience long-range van der Waals interactions which are visualized by the black lines (shown only for the atom at the position $\vec{R}_1$).
The opacity of the lines reflects the interaction strength between a pair of atoms. (b) Level scheme of a single Rydberg atom: The ground state $\ket{g}$ is coupled with the Rydberg state $\ket{r}$ with a coupling $\Omega(t)$ and detuning $\Delta(t)$.}
\label{fig:ring_geometry}
\end{figure}

The associated time-dependent system Hamiltonian reads in natural units ($\hbar = 1$) 
\begin{align}
H(t)= \Omega(t)\, H_{\Omega} - \Delta(t)\, H_{\Delta} + H_d,
\label{eqn:resource_ham}
\end{align}
where $\Omega(t)$ and $\Delta(t)$ denotes the time-dependent Rabi frequency and detuning of a globally acting laser field, respectively. 
The associated coupling Hamiltonians are defined as $H_{\Omega} := \sum_{n=1}^N \sigma^{(n)}_x$ and $H_{\Delta} := \sum_{n=1}^N \sigma^{(n)}_z$, where $\sigma^{(n)}_j$
is a Pauli matrix $\sigma_j$ with $j\in\{x,y,z\}$ acting on the $n^{\mathrm{th}}$ atom.
We are using the notation
\begin{equation*}
\sigma_x 
:=
\left(
\begin{smallmatrix}
0 & 1\\
1 & 0
\end{smallmatrix}
\right),\,
\sigma_y
:=
\left(
\begin{smallmatrix}
0 & -i\\
i & 0
\end{smallmatrix}
\right),\,
\sigma_z 
:=
\left(
\begin{smallmatrix}
1 & 0\\
0 & -1
\end{smallmatrix}
\right),\,
\unity =
\left(
\begin{smallmatrix}
1 & 0\\
0 & 1
\end{smallmatrix}
\right).
\end{equation*}
The term $H_d$, called drift Hamiltonian, describes the interaction between atoms in the Rydberg state and is in general given by
\begin{align}\label{eqn:ring_drift_general} 
    H_d := \sum_{1\leq n<m \leq N}\frac{C_6}{|\vec{R}_n{-}\vec{R}_m|^6}(\unity{-}\sigma^{(n)}_z)(\unity{-}\sigma^{(m)}_z).
\end{align}
The strength of interaction is determined by the $C_6$-coefficient of the considered Rydberg state and the distance between the pair of atoms at positions $\vec{R}_n$ and $\vec{R}_m$.

For the remainder of this manuscript, we consider the $N$ atoms to be arranged equidistantly on a ring as shown in Fig.~\ref{fig:ring_geometry}. 
Due to the rotational symmetry of the atom arrangement the single particle terms of $H_d$ from Eq.~\eqref{eqn:ring_drift_general} can be absorbed in the global detuning $\Delta(t)$ and $H_d$ can be written as
\begin{equation}\label{eqn:ring_drift} 
H_d:=\sum_{1\leq n<m \leq N}\frac{C_6}{|\vec{R}_n{-}\vec{R}_m|^6}\sigma^{(n)}_z\sigma^{(m)}_z,
\end{equation}
up to a global shift. 
In the ring geometry neighboring atoms experience the strongest interaction strength ($C_6/|\vec{R}_n{-}\vec{R}_{n\pm 1}|^6$) and atoms sitting on opposite sides of the ring are interacting weakest.

\subsection{VQE and the target Hamiltonian}

\begin{figure}
\includegraphics{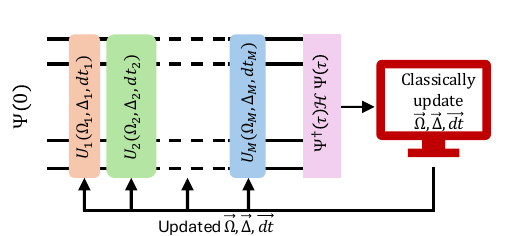}
\caption{VQE circuit executed on the analog Rydberg-atom device by applying parametrized unitaries $U_j$ for 
durations $\delta t_j$. The unitaries $U_j$ are generated by globally acting controls $\Omega_j$ and $\Delta_j$
[see Eq.~\eqref{eqn:resource_ham}]. The expectation value of the target Hamiltonian
$\mathcal{H}$ with final state state $\Psi(\tau)$ is minimized by classically updating the VQE parameters
$\vec{\Omega}$, $\vec{\Delta}$ and $\vec{\delta t}$. }
\label{fig:circuit}
\end{figure}

The hybrid quantum-classical VQE algorithm prepares trial states $\Psi(\tau)$ that aim to approximate the ground state of a given target Hamiltonian $\target$.
The VQE circuit is composed out of $M$ layers of unitaries $U_j$, with $j\in\{1,\,\dots,\,M\}$. Each unitary $U_j$ is generated by time evolution of a set of resource Hamiltonians with associated external control knobs that act as variational parameters. 
The set of resource Hamiltonians considered throughout this work is constructed from the different drift and driving Hamiltonians following the hardware constraints
detailed in Sec.~\ref{sec:VQE:model} and is given by $\gens_R := \{H_d, H_{\Omega}, H_{\Delta}\}$.
The external control (variational) parameters are the Rabi frequency $\Omega_j$, the detuning $\Delta_j$, and the evolution duration $\delta t_j$ of each layer $j$.
Note, the parameters $\Omega_j$ and $\Delta_j$ are considered to be constant over the time interval $\delta t_j$.
The trial state $\Psi(\tau)$ is prepared by evolving an initial state $\Psi(0)$ with the VQE circuit
\begin{equation*}
\Psi(\tau)=U_M(\Omega_M,\Delta_M,\delta t_M) \cdots U_1(\Omega_1,\Delta_1,\delta t_1)\Psi(0),
\end{equation*}
see Fig.~\ref{fig:circuit}. Here $\tau=\sum_{j=1}^M\delta t_j$ is the total evolution time.

The trial state is used to measure the energy expectation value of the target Hamiltonian $\target$
\begin{align*}
E(\tau)={\Psi^\dagger(\tau)}\, \target\, {\Psi(\tau)}.
\end{align*}
This energy $E(\tau)$ defines the cost function and is minimized by a classical optimizer. 
The classical optimization iteratively updates the variational parameters of the VQE circuit: $\vec{\Omega}=(\Omega_1,\cdots,\Omega_M)$,
$\vec{\Delta}=(\Delta_1,\cdots,\Delta_M)$, and $\vec{\delta t}=(\delta t_1,\cdots,\delta t_M)$.
After several iterations, the VQE circuit prepares a state with an energy close to the ground state of
the target Hamiltonian $\target$.

The success of VQE optimization relies heavily on the chosen ansatz and the initial state. 
It is generally difficult to predict whether a certain set of quantum resources is able to prepare the ground state of a particular target Hamiltonian $\target$.
We examine this question in the succeeding sections for the groundstate of the Ising and the Heisenberg Hamiltonian on the ring illustrated in Fig.~\ref{fig:ring_geometry}.

\begin{figure}
\includegraphics{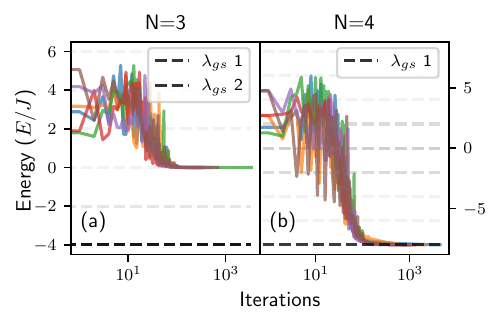}
\caption{VQE for the Heisenberg model: (a)-(b) Variational energy with 10 random circuit initializations as a function of the optimization iterations ($5000$ iterations per initialization) for three and four sites. The energy of the variational states is compared to the eigenenergies  $\lambda$ of the target Hamiltonian $\target_H$ (grey dashed lines) and the ground state energy $\lambda_{gs}$ (black dashed line) for $J=h$.
(a)~The three-site system has two degenerate ground states. VQE does not reach the ground state and gets stuck in the second excited state.
(b)~VQE reaches the ground state of $\target_H$ for four sites in all ten runs.
\label{fig:VQE_heisenberg_3_4}}
\end{figure}

The Ising model's Hamiltonian is in the presence of a longitudinal and transverse field given by
\begin{align}
\target_{I}&={\Omega}\sum_{n=1}^N \sigma^{(n)}_x -\Delta\sum_{n=1}^N \sigma^{(n)}_z
+J\sum_{n=1}^N \sigma^{(n)}_z\sigma^{(n+1)}_z.
\label{eqn:ising}
\end{align}
This model resembles the Hamiltonian of the Rydberg-atom simulator in Eq.~\eqref{eqn:resource_ham} except for the drift term, which is truncated to nearest neighbor. 
The target Hamiltonian of the Heisenberg model is given by
\begin{align}
\target_{H}=h\sum_{n=1}^N \sigma^{(n)}_z +J\sum_{n=1}^N \sum_{j\in \{x,y,z\}}\sigma_j^{(n)}\sigma_j^{(n+1)}.
\label{eqn:heisenberg}
\end{align}
Here, $J$ denotes the isotropic interaction strength between neighboring spins, and $h$ denotes the strength of an  effective magnetic field in $z-$direction.
Preparing the groundstate of these Hamiltonians presents a promising avenue for research. 
For example, the Ising model on a 2D triangular lattice with $J < 0$ exhibits frustration \cite{andre_frustration_1979}, while the Heisenberg model on a 2D triangular lattice has been studied in the context of high-temperature superconductivity and the quantum Hall effect \cite{lima_magnon_2019}.

In our reachability analysis, we consider for the Ising Hamiltonian $\target_{I}$ of Eq.~\eqref{eqn:ising} the parameter choice $\Omega=\Delta=J$.
Similarly, for the Heisenberg Hamiltonian from Eq.~\eqref{eqn:heisenberg}, we set $h=J$.
As a starting point, we discuss here VQE of the Heisenberg target Hamiltonian $\target_{H}$ for three and four sites. 
Figure~\ref{fig:VQE_heisenberg_3_4} presents optimization trajectories, i.e.\ trial state energy as a function of optimization steps. 
Each curve represents one of ten optimization runs,
all starting from the same inital state but using different random starting points in the parameter space.
Figure~\ref{fig:VQE_heisenberg_3_4}(a) illustrates that the VQE is unable to reach the ground state of $\target_{H}$
for three sites and gets stuck in one of the excited states. 
But, for four sites, the VQE reaches the ground state of $\target_{H}$ in all ten runs [see Fig.~\ref{fig:VQE_heisenberg_3_4}(b)].
The difference between the three and four-qubit case suggests a symmetry obstruction for reaching the ground state.

\section{Tools for analyzing reachability\label{sec:tools}}

We aim at developing criteria to decide if the ground state(s)
of a target Hamiltonian $\target$ are reachable from the initial state
by following dynamics facilitated by the resource Hamiltonians
$\gens_R := \{H_d, H_{\Omega}, H_{\Delta}\}$.
In order to understand the restrictions for VQE with a Rydberg-atom simulator, we consider multiple
tools, some of which will be more useful and effective than others.

In Sec.~\ref{lie_algebra}, we recall the notion of a Lie algebra and its reductive decomposition (vide infra).
Determining the Lie algebra $\g_{R}:=\lie{\gens_R}$ 
generated by the resource Hamiltonians
$\gens_R$ and the reductive decomposition of $\g_{R}$ can provide valuable information
about the reachability and will help us to understand the structure of few-qubit examples.
Indeed, combining the knowledge of the Lie algebra and its reductive decomposition
with the structure of the invariant subspaces (see Sec.~\ref{symmetry_analysis})
provides a complete picture to decide reachability. However, computing the Lie algebra
will be in larger cases quite challenging, and even more so for the reductive decomposition.
Therefore, we would like to avoid the computation of the Lie algebra in larger examples.

Much more computationally feasible is the symmetry analysis in Sec.~\ref{symmetry_analysis}
based on the invariant subspaces, especially
if we restrict us to the coarse-grained structure determined by the isotypic subspaces (vide infra).
Thus one proposed approach to the reachability relies on the invariant
subspaces, which leads to necessary conditions depending on
whether the initial and target ground states lie in the same isotypic (or irreducible) subspace.
Even beyond direct computations,
partial knowledge of the inherent symmetries of the resource Hamiltonians $\gens_R$
based on physical considerations can already yield obstructions to the reachability of the
ground state(s) of the target Hamiltonian $\target$. Moreover, we
can potentially extrapolate from smaller examples to provide information
on the reachability of larger examples. In the later Sec.~\ref{sec:adiabatic}, we propose a complementary approach to reachability based on
adiabatic state preparation and we connect this approach to the structure of the invariant subspaces.

Before the symmetry analysis in Sec.~\ref{symmetry_analysis},
we study in Sec.~\ref{simulability} to what degree the simulability (vide infra) of the target Hamiltonian
$\target$ relates to the reachability of its ground states. This does neither 
lead to necessary nor sufficient conditions, but we consider it important to clarify and emphasize the distinction
between simulability and reachability.

\subsection{Lie algebras\label{lie_algebra}}

We start by recalling
Lie algebras and their reductive decompositions \cite{hall2015,jacobson1979,deGraaf2000}
generated by the resource Hamiltonians
$\gens_R$ and similarly when extending the generators with either the target Hamiltonian
$\target_I$ or $\target_H$ [see Eqs.~\eqref{eqn:ising} and \eqref{eqn:heisenberg}]. 
Recall that the (real) Lie algebra $\g:=\lie{\gens}$ generated by a set of hermitian Hamiltonians $\gens=\{H_1,....,H_p\}$ (such as $\gens_R$) 
contains all (real) linear combinations $r_j h_j{+}r_k h_k$ with $r_j \in \R$ and all commutators $[h_j,h_k]:=h_j h_k{-}h_k h_j$ for each $h_j, h_k  \in \g$, while starting from the skew-hermitian elements $h_j := iH_j$.
The dimension of $\g$ is given by its maximal number of linearly independent elements.  

The reductive decomposition $\g= \g_1\rd \cdots \rd \g_k$ decomposes  $\g$ into its simple or abelian components $\g_j$
with $[\g_j, \g_k]=0$. In this work, the symbol 
$\rd$ refers to direct sums of simple or abelian components of a Lie algebra.
In particular, the center $Z\mathfrak{(g)}$ of a Lie algebra $\g$ consists of all its elements $z \in \g$ such that $[z, h_j] = 0$ for all $h_j \in \g$.
An $m$-dimensional center $Z\mathfrak{(g)}$ will be represented by a direct sum $\uu(1)\oplus\cdots\oplus\uu(1)$ of $m$ one-dimensional
abelian components. The corresponding computational techniques are detailed in Appendix~\ref{app:lie_algebra}.

\subsection{Simulability analysis\label{simulability}}

As a first attempt to study the reachability, we analyze the simulability 
of the target Hamiltonian $\target$ by applying the resource Hamiltonians
based on computing the relevant Lie algebras (see Sec.~\ref{lie_algebra}).
We discuss whether this
will resolve the reachability question.
But, below, we will 
conclude that this is not the case!

The set of resource Hamiltonians $\gens_R$ can simulate the dynamics of the target Hamiltonian $\target$ if and only if the Lie algebra
$\h:=\lie{\gens_R \cup \{\target\}}$ is contained in the Lie algebra $\g_{R}:=\lie{\gens_R}$,
or equivalently, iff $\g_R {=}\h$ \cite{jurdjevic_control_1972,zimboras_symmetry_2015}.
Note that $\h$ is generated by $\target$ and the resource Hamiltonians $\gens_R$,
while $\g_{R}$ is generated only by the resource Hamiltonians $\gens_R$.

We consider the Heisenberg target Hamiltonian and
determine the Lie algebras $\g_R$ and $\h_{H}:=\lie{\gens_R \cup \{\target_H\}}$
for three and four qubits. The corresponding Lie algebras are shown in Table~\ref{table:3_4_heisen}
by detailing their reductive decompositions and centers.  We observe that the Lie algebras $\g_R$ and $\h_H$ agree for three qubits but differ by an element of the center in $\h_H$ for four qubits. The corresponding dimensions of the center of $\g_R$ and $\h_H$  are one and two for four qubits, whereas they are the same for the three-qubit system. The simulability condition of $\h_H = \g_R$ suggests that the VQE, in principle, can simulate the Heisenberg Hamiltonian for a system of three qubits. However, simulating the dynamics of $\target_H$ by resource Hamiltonians is not completely feasible for a system of four qubits (due to the additional center element).

\begin{table}
\centering
\begin{tabular}{l@{\hspace{4mm}} l@{\hspace{-2mm}} r}
\hline\hline
\\[-2mm]
Number& Resource Lie & Extension by adding the\\
 of qubits & algebra $\g_R$ & Heisenberg target $\target_H$
\\[1mm]
\hline 
\\[-2mm]
 3 & $\mathfrak{su}(4) {\rd} \mathfrak{su}(2){\rd} \BLUE{\uu(1)} $ & -- \\[1mm]
4 & $\mathfrak{su}(6) {\rd} \mathfrak{su}(3){\rd} \mathfrak{su}(3) {\rd} \BLUE{\uu(1)}$& $\rd \BLUE{\uu(1)}$ 
\\[1.5mm]
\hline\hline
\end{tabular}
\caption{Lie algebras and simulability  analysis without and with the Heisenberg target Hamiltonian. We compare the Lie algebra $\g_R$ of the infinitesimal resource Hamiltonians $\gens_R$ with the Lie algebra $\h_H$ of the resource Hamiltonians combined with the Heisenberg target Hamiltonian, i.e. $\gens_R \cup \{\target_H\}$, for three and four qubits. We describe the Lie algebras by their reductive decompositions.
The dimension of the center $Z\mathfrak{(g)}$ is equal to the number of one-dimensional
abelian components $\uu(1)$. For the three-qubit case, the Lie algebras $\g_R$ and $\h_H$
agree (and their centers also agree). For four qubits,
the dimension of $\h_H$ is by one larger than $\g_R$ due to an additional center element.
The additional center element in the extended Lie algebra for four qubits
is acting on all $\C^3$ blocks in Fig.~\ref{3_4_heisenberg_invariant}.
 \label{table:3_4_heisen}}
\end{table}

How does this compare with the VQE results presented in Fig.~\ref{fig:VQE_heisenberg_3_4}?
Contrary to the suggestions from the simulability analysis, the VQE was unable to reach the
ground state energy for the three-qubit system and 
successfully converged to the ground state for the four-qubit system. 
These observations imply that the simulability analysis alone is not enough to capture the symmetry constraints governing VQE optimization. Consequently, we need to provide an alternative approach
that could (partially) predict the reachability of the ground state of the target Hamiltonian.

 \subsection{Symmetry analysis and invariant subspaces\label{symmetry_analysis}}
We study the symmetries and invariant subspaces to assess the reachability the target ground state(s)
for the given resource Hamiltonians $\gens_R$.
 An \emph{invariant} subspace of $\gens_R$ (and of its generated Lie algebra $\g_R$)
 is a subspace that its mapped to itself by the infinitesimal action of the resource Hamiltonians $\gens_R$.
We refer to    
\cite{lorenz2008,jacobson1985,jacobson1989,
lux2010,curtisreiner1962,curtisreiner1981}
for more general context and results on invariant subspaces and representation theory.

To determine the decomposition into invariant subspaces,
we first compute the commutant of the resource Hamiltonians $\gens_R$ \cite{zimboras_symmetry_2015,zeier_symmetry_2011}.
The \emph{symmetries} of the resource Hamiltonians (and the generated Lie algebra) are formally described by
the commutant.
The \emph{commutant} is the matrix subspace of all complex matrices simultaneously commuting with the Hamiltonians
$\gens_R$ (i.e.\ $[S, iH_v]=0$ for $S \in  \C^{d \times d} $, $d=2^N$, and $H_v \in \gens_R$). It can be calculated as the
simultaneous null space of the matrices $\unity_{d} \otimes H_v-H_v\otimes \unity_{d}$ for all $H_v \in \gens_R$.
This enables us to also calculate the center $\Z(\g_R)$ of the generated Lie algebra from the commutant
(see Appendix~\ref{app:symmetry_analysis}) without determining the Lie algebra $\g_R$ and its reductive
decomposition as in Sec.~\ref{lie_algebra}.

\begin{figure}
\includegraphics{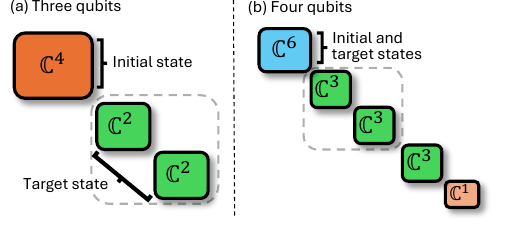}
\caption{Block structure and invariant subspaces: Symmetry analysis of the VQE for the Heisenberg Hamiltonian with three and four qubits showing the 
invariant subspaces and the support  of the initial state and the Heisenberg target ground state. 
Irreducible subspaces inside a gray dashed box are constructed from the same isotypic subspace.
(a)~For three qubits, the two isotypic subspces are
$\C^4$ and $\C^2\ds \C^2$ where the first one $\C^4$ is irreducible and the second
one splits into two irreducible subspaces $\C^2 \ds \C^2$.
Here, $\ds$ represents a direct sum of subspaces. The initial state has only
support in $\C^4$ and the target ground state has only support in $\C^2\ds \C^2$. Clearly, the target ground state
can never be reached.
(b)~For four qubits, both the initial and the target state lie in the same irreducible component $\C^6$.
\label{3_4_heisenberg_invariant}}
\end{figure}

\begin{figure*}
\includegraphics{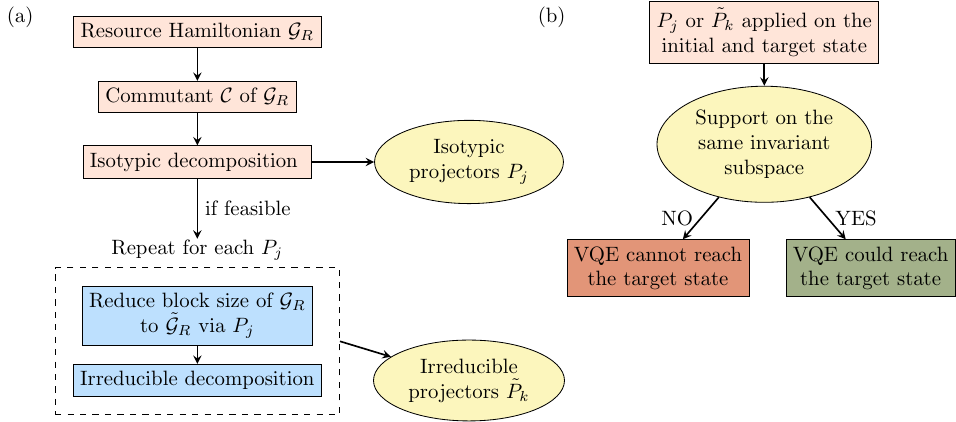}
\caption{Flowchart for the symmetry analysis. (a) First, the commutant $\mathcal{C}$ 
is calculated from the resource Hamiltonians $\gens_R$, followed by the computation of the isotypic projectors $P_j$.
For smaller number of qubits, we can further break up the isotypic
subspace, leading to the irreducible projectors $\tilde{P}_k$. They are obtained by reducing the
resource Hamiltonians to a block size corresponding
to the isotypic subspace of $P_j$. 
(b) The isotypic or irreducible projectors $P_j$ or $\tilde{P}_k$ are applied to the initial and target states.
If the initial and target states belong to the same invariant subspace(s), then the target state could be reached.
\label{fig:overview}}
\end{figure*}

As our action via $\gens_R$ corresponds to a subgroup of the unitary group
(i.e.\ a compact Lie group), we know that
all representations are completely reducible (i.e.\ semisimple) and thus decompose into 
irreducible (i.e.\ simple) representations \cite{hall2015} (and the same applies
to invariant subspaces).
Recall that an invariant subspace is \emph{irreducible}
if it cannot be further decomposed into smaller invariant subspaces.
We denote an invariant subspace
as \emph{isotypic} (or primary) if it contains all and only the irreducible subspaces 
from the decomposition of the whole space
that are isomorphic to each other
(see, e.g., \cite[p.~156]{procesi2007} and \cite[p.~VIII.61]{bourbaki2022}).
We can construct projection matrices to the isotypic
subspaces using effective algorithms based on first computing the center of the commutant
\cite{eberly_efficient_1996}.
If computationally feasible, this is combined with 
the so-called meataxe algorithm
\cite{lux2010} to determine the irreducible subspaces in each isotypic one. But we first
reduce the block size of the resource Hamiltonians by
projecting them to a block size corresponding
to the isotypic components with the help of the isotypic projectors (see App.~\ref{app:symmetry_analysis}). 
Figure~\ref{fig:overview}(a) provides an overview
and App.~\ref{app:symmetry_analysis} details the computation of the isotypic and irreducible projectors.

\begin{figure*}
\includegraphics[scale=0.85]{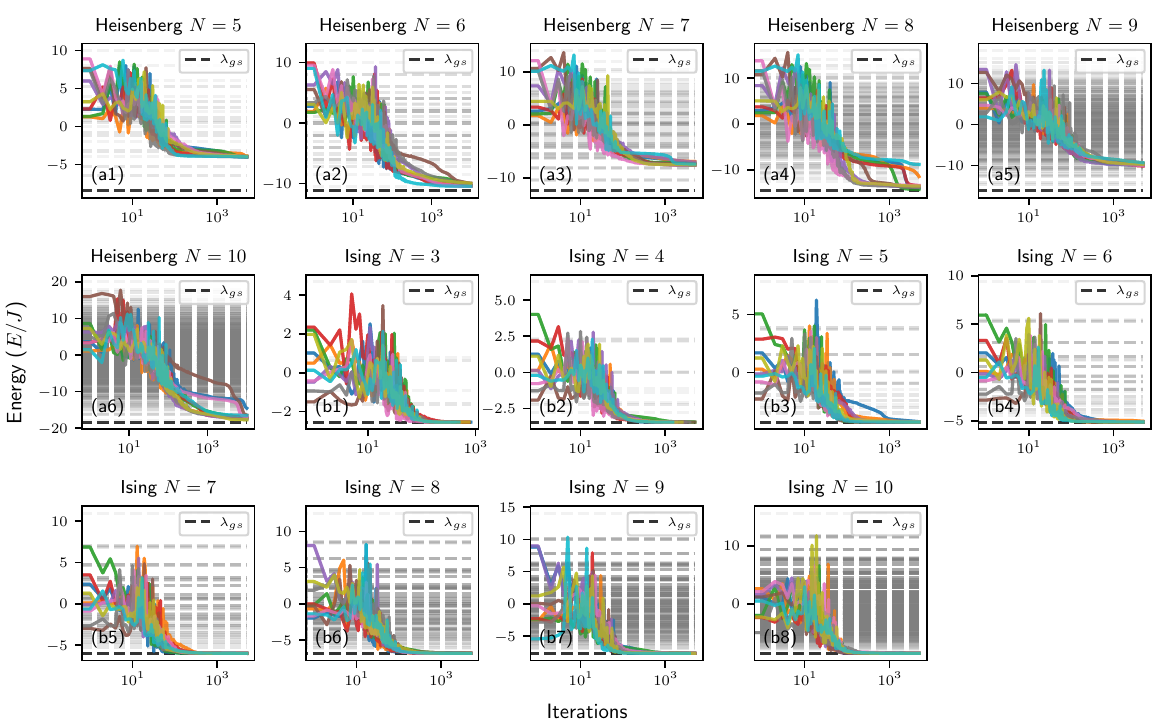}
\caption{VQE for the Heisenberg and Ising ground state showing the expectation value of the energy compared to
the eigenenergies  $\lambda$  (grey horizontal lines) and the ground state energy $\lambda_{gs}$ (black horizontal line)
of the target Hamiltonian as a function of the optimization steps for different number of qubits $N$
and ten random circuit initializations.
(a1)-(a6) For the Heisenberg target Hamiltonian $\target_H$ of Eq.~\eqref{eqn:heisenberg} with $J=h=1$,
the optimization fails to reach the ground state while reaching only an excited state for all qubit numbers, except for $N \in \{4, 8,10\}$. 
(b1)-(b8) For the Ising Hamiltonian $ \target_I$ of Eq.~\eqref{eqn:ising} with $ J=\Omega=\Delta=1$ at least one instance of the optimization runs reaches the ground state energy for all considered qubit numbers,
indicating that the resource Hamiltonians are in principle capable of preparing the ground state.
\label{fig:VQE_mixed}}
\end{figure*}

We now discuss how the structure of the invariant subspaces affects the reachability of the ground state(s) from the initial state
during the execution of VQE  by providing necessary conditions. For this discussion, we assume that we know the ground
state(s) by explicitly calculating them for the target Hamiltonian. This will not be possible for high qubit numbers but allows to
get some insight into this structure starting from a small number of qubits. We project the computed ground state(s) with the 
isotypic (or the irreducible) projectors. Similarly, the conventional initial state $(1,0,\ldots,0)^T\in \C^d$ for $d=2^N$ is projected.
We record the supports in the various invariant subspaces, i.e., whether the projection is not zero. 
Clearly, any infinitesimal action of the resource Hamiltonians $\gens_R$ on the initial state with support in one or more
multiple invariant subspace(s) will preserve these invariant subspace(s). Thus the VQE can only reach
the ground state(s) from a particular initial state if they lie in the same invariant subspace(s). 

However, the condition that the initial and the target state lie
in the same (irreducible) invariant subspace is not 
sufficient for the target state to be reachable from the initial state.
This can even seen in a simple two-qubit example
\cite{zeier_symmetry_2011,zimboras_symmetry_2015}: both local unitary operations
$\SU(2)\otimes\SU(2)$ and the full special unitary group $\SU(4)$ act irreducibly on
the four-dimensional Hilbert space, while  local unitary operations \emph{cannot}
reach entangled states from non-entangled ones. Thus our analysis based on containment
in the same invariant subspace(s) will provide only a necessary condition for reachability.

We will now illustrate this analysis with the Heisenberg example for three and four qubits.
Figure~\ref{3_4_heisenberg_invariant} shows the support for the initial and the target state on the occurring invariant subspaces.  
For three qubits, one observes the three irreducible, invariant subspaces $\C^4 \ds \C^2 \ds\C^2$, where both
$\C^4$ and $\C^2\ds \C^2$ are isotypic subspaces. 
Here, $\ds$ represents a direct sum of subspaces.
The initial state has only support in the first isotypic subspace
and the target state has only support in the second isotypic subspace.
This implies that the VQE can never reach the Heisenberg target ground state for three qubits while starting from our choice of initial state. 
For four qubits, one obtains the invariant subspaces $\C^6\ds \C^3\ds \C^3\ds \C^3\ds \C^1$ with the four isotypic subspaces
$\C^6$, $\C^3$, $\C^3\ds \C^3$, and $\C^1$. Both the initial and the target state have a nonzero support only in the first isotypic (and irreducible) subspace $\C^6$. Thus the necessary condition for reachability is fulfilled and the symmetries of the resource Hamiltonians $\gens_R=\{H_d, H_\Delta, H_\Omega\}$ do not prevent us from reaching the Heisenberg target ground state.
This analysis is consistent with the VQE simulations in Fig.~\ref{fig:VQE_heisenberg_3_4}.

We emphasize that adding the Hamiltonian $\target_H$ to the resources would not help in reaching the target state.
For the discussed three-qubit case, the generated Lie algebra $\g_R$ is unchanged (see
Table~\ref{table:3_4_heisen}). Although the Lie algebra $\g_R$ is extended by an additional center element
for four qubits, the support of this element is restricted to outside of the subspace $\C^6$ which contains
both the initial and the target state [as visualized in Fig.~\ref{3_4_heisenberg_invariant}(b)]. Consequently, this
additional center element cannot affect the reachability of the Heisenberg target ground state.

\section{Larger system sizes\label{sec:larger_system}}

The VQE optimizations shown in Fig.~\ref{fig:VQE_heisenberg_3_4} aim at finding the ground state of the Heisenberg Hamiltonian of Eq.~\eqref{eqn:heisenberg}, they are now extended for up to ten qubits in Fig.~\ref{fig:VQE_mixed}(a1)-(a6).
The VQE only succeed for four, eight, and ten qubits.
Whereas, Fig.~\ref{fig:VQE_mixed}(b1)-(b8) reveals that the VQE optimizations
are successful for 
the Ising Hamiltonian of Eq.~\eqref{eqn:ising} with three to ten qubits. 

\begin{table}
\centering
\begin{tabular}{l@{\hspace{5mm}} c @{\hspace{4mm}} c @{\hspace{4mm}} c }
\hline\hline
\\[-2mm]
& \multicolumn{3}{c}{Support} \\[0.5mm]
\cline{2-4}
\\[-2.5mm]
Number
&&Heisenberg& Ising\\
of qubits & Initial state & target state(s)& target state(s)
\\[1mm]
\hline
\\[-2mm]
3 &$\RED{\C^4}$&$(\BLUE{\C^2}{\ds} \GREEN{\C^2})$ &$\RED{\C^4}$
\\[1mm]
4 & $\RED{\C^6}$&$\RED{\C^6}$&$\RED{\C^6}$
\\[1mm]
5 &$\RED{\C^8}$&$(\BLUE{\C^6}{\ds} \GREEN{\C^6})$&$\RED{\C^8}$
\\[1mm]
6 &  $\RED{\C^{13}}$&$\BLUE{\C^{7}}$&$\RED{\C^{13}}$
\\[1mm]
7 & $\BLUE{\C^{18}}$&$(\GREEN{\C^{18}}{\ds} \VIOLET{\C^{18}})$& $\BLUE{\C^{18}}$
\\[1mm]
8 &$\MAG{\C^{30}}$&$\MAG{\C^{30}}$&$\MAG{\C^{30}}$
\\[1mm]
9&$\GREEN{\C^{46}}$&$(\VIOLET{\C^{56}}{\ds} \VIOLET{\C^{56}})$&$\GREEN{\C^{46}}$
\\[1mm]
10 &$\MAG{\C^{30}}$&$\MAG{\C^{30}}$&$\MAG{\C^{30}}$
\\[1.5mm]
\hline\hline
\end{tabular}
\caption{Reachability of the Heisenberg and Ising target states for the VQE with the Rydberg-atom platform of Sec.~\ref{sec:VQE}.
The invariant subspaces are determined together with the support of the initial state and the Heisenberg target state for three to ten qubits. 
The full decomposition of the invariant subspaces for the Heisenberg and Ising model is shown in Table~\ref{table_app:symmetry}
in Appendix~\ref{app:symmetry_analysis}. Two or more (isomorphic) irreducible subspaces inside two parentheses belong to
one isotypic subspace. An isotypic subspace decomposes into irreducible subspaces
which are represented with different colors if we have computationally determined the respective irreducible projectors
(which applies only for up to eight qubits).
The initial states for the VQE are same for
both Heisenberg and Ising case.
For the Heisenberg model, the initial and the target states are shown to be in different isotypic subspaces,
except for four, eight, and ten qubits. This implies that the Heisenberg target state can never be reached for three, five, six, seven, and nine qubits using the chosen initial state. 
For the Ising case, the initial and the target states are always in the same irreducible subspace for three to ten qubits.
\label{table:symmetry}}
\end{table} 

This is now compared with the corresponding results from the symmetry analysis
as summarized in Table~\ref{table:symmetry}. Following the analysis in Sec.~\ref{symmetry_analysis},
Table~\ref{table:symmetry} lists the non-zero support of the initial state as well as the
target ground state(s). We observe for five-qubit
Heisenberg case that the initial state lies in the invariant subspace $\C^8$, while
the ground state of the Heisenberg Hamiltonian lies in one of the isotypic subspaces
of the form $\C^6 \ds \C^6$. Thus the ground state cannot be reach by the VQE.
A similar analysis 
rules out the reachability for six, seven, and nine qubits in agreement with the VQE results
from Fig.~\ref{fig:VQE_mixed}(a1)-(a6). For four, eight, and ten qubits, the initial and the target ground
state are contained in the same irreducible subspace which corresponds to the successful VQE runs in Fig.~\ref{fig:VQE_heisenberg_3_4}(b) and 
Fig.~\ref{fig:VQE_mixed}(a1)-(a6).
On the other hand,
the symmetries of the resource Hamiltonians do not prevent us from reaching the Ising target ground state during the VQE as they both only have a nonzero support in a same irreducible subspace for up to ten qubits.
This agrees with the corresponding VQE optimizations in Fig.~\ref{fig:VQE_mixed}(b1)-(b8).

Similar to the discussion in Sec.~\ref{symmetry_analysis}, we check whether adding
the target Hamiltonian would affect and change the result of the reachability analysis in Appendix~\ref{app:lie-center}.

\section{Initial state selection\label{sec:initial_state}}

One straightforward approach to resolve the symmetry restrictions resulting from the resource Hamiltonians is to choose a different initial state.
The trivial initial state considered in the manuscript so far is the unique groundstate of $-\sum_{n=1}^N\sigma_{z}^{(n)}$, i.e.\ the state vector
 $(1,0,\ldots,0)^T\in \C^d$ for $d=2^N$ in the conventional computational basis ordering.
As clarified by the symmetry analysis, VQE can successfully prepare the ground state only if the initial state and the target ground state are situated within the same invariant subspace (which can be either an isotypic or irreducible subspace).
Thus, any state residing in a suitable invariant subspace containing the target state can serve as an effective initial state for VQE.

\begin{figure}
\includegraphics{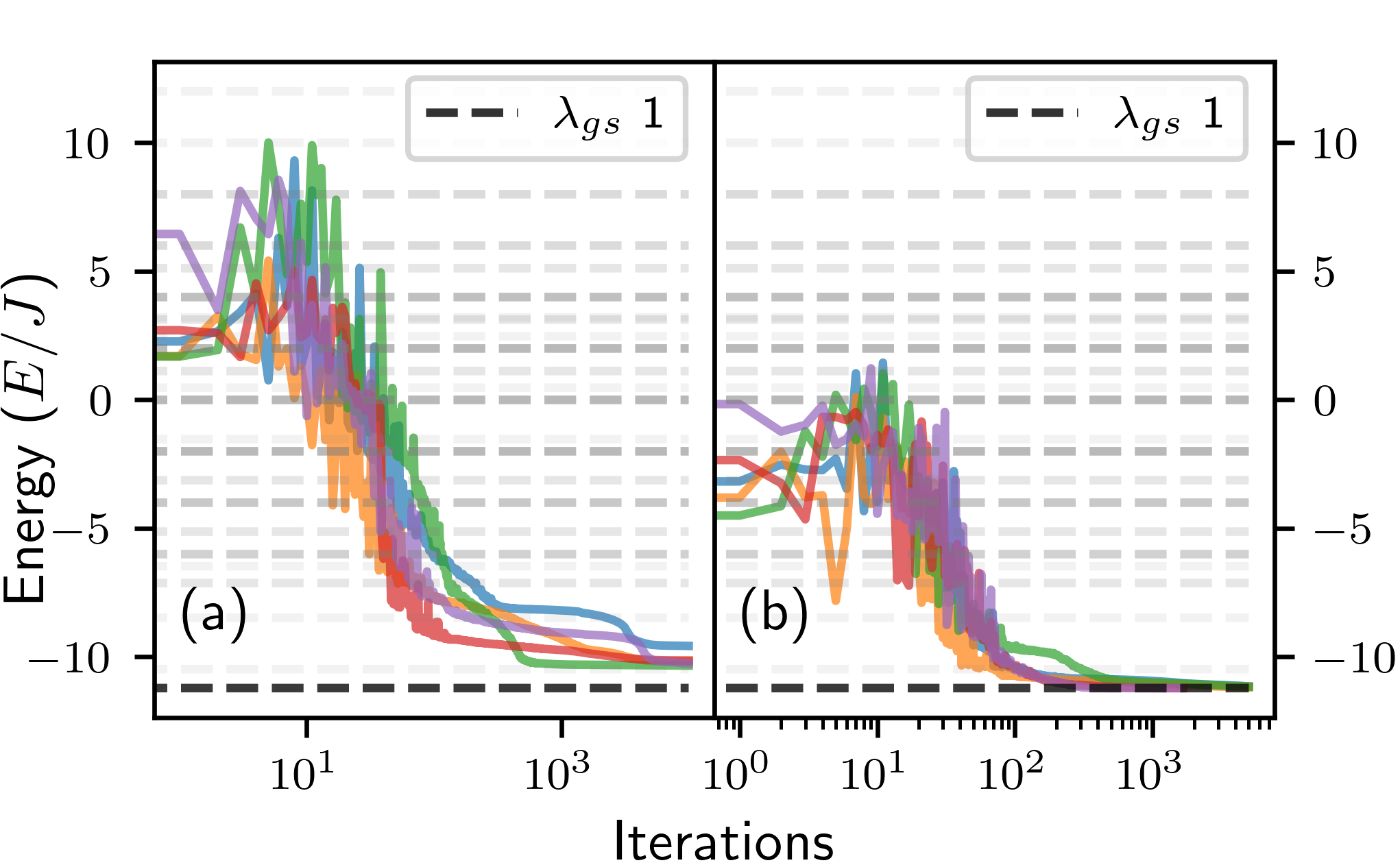}
\caption{Comparison of VQE performance with different initial states for the six-qubit Heisenberg target ground state.
(a)~The VQE fails to converge to the target ground state when initialized from the trivial initial state introduced in Sec.~\ref{sec:initial_state} using six layers and 10000 iterations, indicating confinement within the $\C^{13}$ subspace. (b) By initializing VQE with
an eigenvector $\nu_i$ from the projection operator related to the $\C^{7}$ subspace, the ground state is 
reached in all ten optimization runs in 5000 iterations.}
\label{fig:initial_state_prep}
\end{figure}

We consider the six-qubit Heisenberg case and compute the eigenvectors of the projector 
for the invariant subspace associated with the target ground state.
All these eigenvectors, along with their linear combinations, could be suitable initial states for the VQE process.
As illustrated in Fig.~\ref{fig:initial_state_prep}(a), the VQE fails to converge to the target ground state when initialized from the initial state introduced earlier,
even with six layers and after $10^4$ iterations.
The VQE is confined to the  subspace $\C^{13}$ and cannot access the ground state (see Table~\ref{table:symmetry}).
To resolve this, we compute the projection operator associated with the subspace $\C^{7}$
connected to the target state
and determine its eigenvectors
$\nu_j$. We run the VQE while using one of the eigenvectors $\nu_j$ as initial state.
The results of this approach are presented in Fig.~\ref{fig:initial_state_prep}(b) and
one observes a fast convergence to the ground state.

\section{Connections to adiabatic state preparation\label{sec:adiabatic}}

An interesting parallel can be drawn between variational state preparation and adiabatic state preparation or
quantum annealing~\cite{farhi2000}. Indeed, original proposals for variational state preparation were inspired
by trotterizing adiabatic paths~\cite{farhi_2014_QAOA}, and later works found for instance that variational
parameter initializations mimicking an adiabatic path provided good initial guesses for the parameter optimization process~\cite{Sack_2021}.
In this context, we now relate our symmetry analysis to adiabatic state preparation 
and find that some conclusions
can be carried over between the two protocols. Moreover, insights from adiabatic state preparation 
enable us to establish an adiabatic path from the initial state $\ket{\psi_0}$
to a ground state of the target Hamiltonian $\target_T$ assuming a spectral gap is separating the ground states
from the rest of the spectrum
during the adiabatic evolution (and additional assumptions).
This perspective is connected to the structure of the invariant subspaces
with the help of three illustrative examples.

We start by relating the initial state $\ket{\psi_0}$
to a specific initial parent Hamiltonian $\target_{0}$ that has $\ket{\psi_0}$ as one of its ground states.
Our goal is to reach a ground state of a given target Hamiltonian $\target_T$ by slowly varying the time-dependent
Hamiltonian
\begin{align}
H(\tau) = (1{-}\tau) \target_{0} + \tau \target_{T}.
\label{eqn:H_tau}
\end{align}
Here, the parameter $\tau$ changes from $0$ to $1$ while
$H(0)=\target_{0}$ and $H(1)=\target_{T}$. 
In the adiabatic theorem~\cite{BornFock1928,Kato1950,Messiah1962,teufel2003,Teufel2022,Simon2019},
one usually assumes a nonzero gap between the ground-state energy of $H(\tau)$ and
its second lowest energy for $0 \leq \tau \leq 1$ (among more technically assumptions).
If the the system starts at $\tau=0$ in the ground state of $\target_{0}$ and  the time evolution of the Hamiltonian is sufficiently slow,
the evolved state will then approximately remain in a ground state throughout the evolution, thus approximately resulting in the ground state of $\target_{T}$.

\begin{figure}
\includegraphics{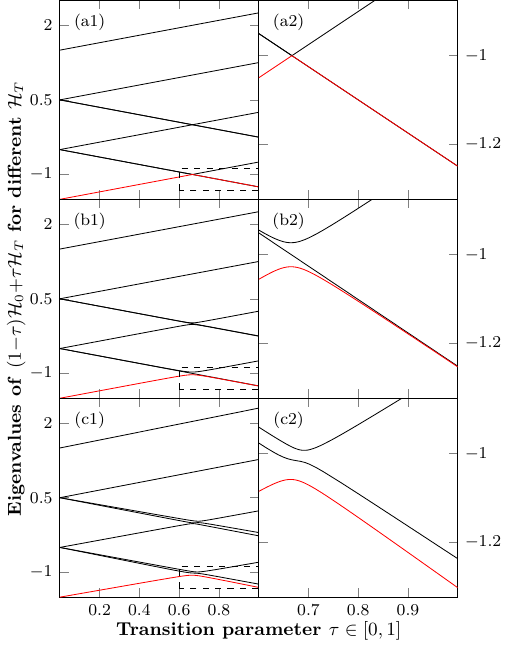}
\caption{Energy spectrum of $H(\tau)$ [see Eq.~\eqref{eqn:H_tau}] with $\tau$ ranging from $0$ to $1$
for the initial parent Hamiltonian $\target_0$ [see Eq.~\eqref{eqn:parent_ham}] and
different target Hamiltonians $\target_T\in \{\target_1, \target_{2}, \target_{3}\}$ [see Eqs.~\eqref{eqn:H0}-\eqref{eqn:H2}].
The energy spectrum of $H(\tau)$ with $\target_1$ is illustrated in (a1), where (a2) contains 
a magnified version of the
dashed rectangular part in (a1).
The ground state energy is represented using the red solid line. Similarly (b1)-(b2) and (c1)-(c2) correspond to $\target_{2}$ and $\target_{3}$ respectively.
\label{fig:adiabatic}}
\end{figure}

To verify that the ground state maintains a nonzero energy gap from the first excited state throughout the evolution,
one can determine the energy spectrum of $H(\tau)$ for $\tau$ ranging from $0$ to $1$ (see Fig.~\ref{fig:adiabatic}). In particular,
crossings in the spectrum involving the ground state imply a zero gap.
The adiabatic theorem has been extended to cases without a gap \cite{AvronElgart1999},
which highlights that having a nonzero gap is in general not a necessary condition.

The Hamiltonians $\target_0$ and $\target_T$ can be used perform the described adiabatic evolution
by applying the unitary
\begin{align}
    U = \mathcal{T} \exp\left(-i \int_0^1 H(t) \, dt \right),
\end{align}
where $\mathcal{T}$ denotes the time-ordering operator.
Thus  the Lie algebra $\lie{\{\target_0,\! \target_T\}}$ generated by $\target_0$ and $\target_T$
is able to infinitesimally implement $U$. In this section, we will now make the simplifying assumption
that the resource Lie algebra $\mathfrak{g}'_R$ generated by a given set of resource Hamiltonians
can always generated the dynamics of both $\target_0$ and $\target_T$, i.e.,
$\mathfrak{g}'_R$ always contains both $i \target_0$ and $i \target_T$.
Note that this assumption is not statisfied in all possible cases (see Tables~\ref{table:3_4_heisen} and \ref{table:lie_algebra}).
If our assumption is met, the adiabatic evolution between ground states will be reflected
by the fact they have support within the same invariant subspace of $\{\target_0, \target_T\}$ such that 
the Lie algebra $\mathfrak{g}'_R$ can infinitesimally connect them.

To clarify this idea, we study the case of one initial parent Hamiltonian
\begin{equation}
\target_0=\frac{1}{2}\sum_{n=1}^{3}\sigma_z^{(n)},
\label{eqn:parent_ham}
\end{equation}
and three example target Hamiltonians $\target_T$, under the assumption 
that the resource Lie algebra $\mathfrak{g}'_R$ contains both $\target_0$ and $\target_T$.
In the first example, 
$\target_T$ is given by
 \begin{equation}
 \target_{1} = \frac{1}{2} \sum_{n=1}^3 \sigma^{(n)}_z + \frac{1}{4}  \sum_{n=1}^3 \sum_{j \in \{x,y,z\}} \sigma_j^{(n)} \sigma_j^{(n+1)}.
\label{eqn:H0}
 \end{equation}
Figure~\ref{fig:adiabatic}(a) highlights a crossing in the eigenspectrum of $H(\tau)$
related to the ground-state energy. This results in a zero gap and cannot confirm
an adiabatic path to reach the target ground state.
We can also validate this using our symmetry analysis:
The Lie algebra generated by $\target_0$ and $\target_{1}$ is two-dimensional and abelian as
$[\target_0, \target_{1}]=0$.
The corresponding invariant subspaces of $\target_0 \cup \target_{1}$ are
\begin{equation*}
\C^1 \ds \C^1 \ds (\C^1 {\ds} \C^1) \ds \C^1 \ds \C^1 \ds (\C^1 {\ds} \C^1).
\end{equation*}
We calculate the ground states of $\target_0$ and $\target_{1}$ and project them onto the different invariant subspaces.
We observe that the initial ground state has only support on one of the subspaces $\C^1$, while the two degenerate ground states of $\target_{1}$ have only support on one of the subspaces $(\C^1 \ds \C^1)$.
This clearly rules out any path to the target ground state. But with more resources and a larger resource Lie algebra $\mathfrak{g}'_R$, the invariant subspaces will combine and
one can eventually reach the target ground state.

In the second example, we add a local $\sigma_x$ term to $\target_{1}$ so that the new target Hamiltonian is given by
\begin{equation}
\target_{2} = \target_{1} + \frac{1}{20} \sigma_x^{(1)}.
\label{eqn:H1}
\end{equation}
By inspecting the eigenspectrum of $H(\tau)$ in Fig.~\ref{fig:adiabatic}(b), the gap between the ground state and the first excited state is quite small but remains finite, 
thereby allowing for an adiabatic path between the ground states of $\target_0$ and $\target_{2}$.
This agrees with our symmetry analysis.
The dimension of the Lie algebra $\mathfrak{su}(6) \rd \mathfrak{su}(2) \rd \mathfrak{u}(1)$ generated by $\target_0$ and $\target_{2}$ is $39$.
The Hilbert space decomposes into the invariant subspaces
$\C^2 \ds \C^6$, where both the initial and target ground states have only support on the subspace $\C^6$ on which the Lie algebra acts transitively.

Finally, in the third example, we add an extra $\sigma_z$ term to $\target_{2}$ so that the new target Hamiltonian is given by
\begin{align}
\target_{3} =\target_{2} 
+ \frac{1}{20} \sigma_z^{(2)} = \target_{1} + \frac{1}{20} \sigma_x^{(1)} + \frac{1}{20} \sigma_z^{(2)}.
\label{eqn:H2}
\end{align}
The gap in the eigenspectrum of $H(\tau)$ in Fig.~\ref{fig:adiabatic}(c) is clearly visible which again implies an adiabatic path.
The Lie algebra generated by $\target_0$ and $\target_{3}$ is $\mathfrak{su}(8)$ and therefore has a dimension of 63,
which yields a single invariant subspace
$\C^8$, in which both the initial and target ground states lie.
Thus one can reach the ground state of $\target_{3}$ starting from the ground state of $\target_0$.

These examples highlight a connection between our symmetry results and
adiabatic state preparation. Assuming that the resource Lie algebra $\mathfrak{g}'_R$ contains both $i \target_0$ and $i \target_T$
and that there is a nonzero spectral gap (and further technical assumptions), one can establish an adiabatic path
between the ground states of $\target_0$ and $\target_T$. This complements our reachability analysis
based on symmetries. But even 
without an adiabatic path, one reach the target ground state if one increases the available resources 
to enlarge the generated Lie algebra and enable transitions that would otherwise be forbidden.

\section{Conclusion\label{sec:conclusion}}

We have developed symmetry tools to analyze the ground-state reachability for VQE.
These methods have been illustrated and 
applied to a Rydberg-atom platform with restricted global control Hamiltonians.
For a small number of qubits, this enabled us to a priori predict whether the VQE could be successful in reaching
the target state from the chosen initial state. For this, we have analyzed the generated Lie algebras as well
as the invariant subspaces resulting from the restricted control. We have then determined whether
the initial and target states are contained in different invariant subspaces which would rule out the reachability
of the target state. Even beyond a small number of qubits, the developed symmetry tools provide guidance on how inherent
symmetries in controlled quantum systems might limit the applicability of VQE, which could then be translated into
insights into engineering suitable additional interactions in quantum simulation platforms.
Finally, we have connected our symmetry analysis 
to adiabatic state preparation, which can be used
to establish an adiabatic path to the target ground state. This
provides a complementary tool to analyze reachability.

In all our analysis, we have assumed that we have access to the target
ground state (which VQE is supposed to find) or even more
precise spectral information. In a practical, large-scale VQE scenario with restricted controls,
this information will be usually not available. So an open question remains
whether suitable accessible properties of the target Hamiltonian
and, possibly, an initial parent Hamiltonian can be used to confirm reachability.
In absence of such a criteria, more control capabilities
to break inherent symmetries are certainly advantageous.

\begin{acknowledgments}
JS and RZ thank 
Zoltán Zimborás for discussions on symmetries and the adiabatic theorem.
AK and RvB thank
Christian Kokail and Pietro Silvi for illuminating discussions.
RZ appreciates many discussions on symmetries of 
controlled quantum systems with Thomas Schulte-Herbrüggen and
Michael Keyl. Moreover, RZ thanks Roberto Gargiulo, Armin
Römer, Emanuel Malvetti, Sujay Kazi, Martín Larocca, and Marco Cerezo 
for discussions on closely related symmetry topics.
We acknowledge computations with the 
computer algebra system Magma \cite{magma1997}.

The authors acknowledge funding 
under the European High-Performance Computing Joint Undertaking (JU) under grant agreement No 
\href{https://doi.org/10.3030/101018180}{101018180} (HPCQS). 
The JU receives support from the 
European Union’s Horizon 2020 research and innovation programme and Germany, France, Italy, Ireland, Austria, Spain. JS and RZ also acknowledge funding under Horizon Europe programme
HORIZON-CL4-2022-QUANTUM-02-SGA via the project 
\href{https://doi.org/10.3030/101113690}{101113690} (PASQuanS2.1).
JS is also supported by the German Federal Ministry of Education and Research through
the funding program quantum technologies—from basic research to market under the project
MUNIQC-Atoms, \href{https://www.quantentechnologien.de/forschung/foerderung/quantencomputer-demonstrationsaufbauten/muniqc-atoms.html}{13N16073}.
\end{acknowledgments}

\appendix

\section{Lie-algebra computations\label{app:lie_algebra}}

\begin{table}[b]
\centering
\begin{tabular}{l@{\hspace{4mm}} l@{\hspace{2mm}} l@{\hspace{5mm}} r @{\hspace{4mm}} r}
\hline\hline
\\[-2mm]
Number& Resource Lie & \multicolumn{3}{l}{Extension by adding}\\
 of qubits & algebra $\g_R$ && $\target_H$  & $\target_I$
\\[1mm]
\hline 
\\[-2mm]
3 & $\mathfrak{su}(4) {\rd} \mathfrak{su}(2) {\rd} \BLUE{\uu(1)}$ & & -- & -- 
\\[1mm]
4 & $\mathfrak{su}(6) {\rd} \mathfrak{su}(3) {\rd} \mathfrak{su}(3) {\rd} \BLUE{\uu(1)}$
& & $ \rd \BLUE{\uu(1)}$ & $ \rd \BLUE{\uu(1)}$
\\[1mm]
5 & $\mathfrak{su}(8){\rd} \mathfrak{su}(6) {\rd} \mathfrak{su}(6) {\rd} \BLUE{\uu(1)}$ &&--& -- 
\\[1mm]
6 & $\mathfrak{su}(13){\rd} \mathfrak{su}(11) {\rd} \mathfrak{su}(9) $
&&  $ \rd \BLUE{\uu(1)}$& $ \rd \BLUE{\uu(1)}$\\[0mm]
& ${\rd} \mathfrak{su}(7)  {\rd} \mathfrak{su}(3)  {\rd} \BLUE{\uu(1)}$
\\[1mm]
7 & $\mathfrak{su}(18){\rd} \mathfrak{su}(18) {\rd} \mathfrak{su}(18) $
&&  -- & --\\[0mm]
& ${\rd} \mathfrak{su}(18)  {\rd} \mathfrak{su}(2)  {\rd} \BLUE{\uu(1)}$
\\[1mm]
\hline\hline
\end{tabular}
\caption{Lie algebras specified by their reductive decompositions for three to seven qubits: we compare the Lie algebra $\g_R$ of
the infinitesimal resources Hamiltonians $\gens_R$ with the Lie algebras $\h_H$ or $\h_I$ generated by the resources Hamiltonians
combined with the Heisenberg or the Ising target Hamiltonian, i.e.\ $\gens_R \cup \{\target_H$\} and $\gens_R \cup \{\target_I\}$.
For the case of four and six qubits, we observe that $\h_H$ and $\h_I$ have one extra abelian component $\uu(1)$
compared to $\g_R$ and the dimension of the center of the Lie algebra is increased from 1 to 2.
\label{table:lie_algebra}}
\end{table}

\begin{table*}[t]
\centering
\begin{tabular}{l @{\hspace{-4mm}} c @{\hspace{5mm}} c @{\hspace{4mm}} c @{\hspace{4mm}} c}
\hline\hline
\\[-2mm]
&  & \multicolumn{3}{c}{Support}
\\[0.5mm]
\cline{3-5}
\\[-2.5mm]
Number& Invariant & Initial& Heisenberg& Ising\\
 of qubits & subspaces & state & target state(s) &target state(s)
\\[1mm]
\hline
\\[-2mm]
3 &  $\RED{\C^4}{\ds} (\BLUE{\C^2} {\ds} \GREEN{\C^2})$ &$\RED{\C^4}$&($\BLUE{\C^2}{\ds} \GREEN{\C^2}$) &$\RED{\C^4}$
\\[1mm]
4 &  $\RED{\C^6}{\ds} (\BLUE{\C^3} {\ds} \GREEN{\C^3}){\ds} \VIOLET{\C^3}{\ds} \GRAY{\C^1}$&$\RED{\C^6}$&$\RED{\C^6}$&$\RED{\C^6}$
\\[1mm]
5 &  $\RED{\C^8}{\ds} (\BLUE{\C^6} {\ds} \GREEN{\C^6}){\ds} (\VIOLET{\C^6}{\ds} \GRAY{\C^6}$)&$\RED{\C^8}$&$(\BLUE{\C^6}{\ds} \GREEN{\C^6})$&$\RED{\C^8}$
\\[1mm]
6 &  $\RED{\C^{13}}{\ds} (\BLUE{\C^{11}} {\ds} \GREEN{\C^{11}}){\ds} (\GRAY{\C^9}{\ds} \RED{\C^9}){\ds} \BLUE{\C^7}{\ds} \GREEN{\C^3}{\ds} \VIOLET{\C^1}$ &$\RED{\C^{13}}$&$\BLUE{\C^{7}}$&$\RED{\C^{13}}$
\\[1mm]
7 &  $(\RED{\C^{18}}{\ds} \BLUE{\C^{18}}) {\ds} (\GREEN{\C^{18}}{\ds} \VIOLET{\C^{18}}){\ds} (\GRAY{\C^{18}}{\ds} \RED{\C^{18}}){\ds} \MAG{\C^{18}}{\ds} \GREEN{\C^2}$ &$\MAG{\C^{18}}$&$(\GREEN{\C^{18}}{\ds} \VIOLET{\C^{18}})$&$\BLUE{\C^{18}}$
\\[1mm]
8 &  $(\RED{\C^{33}}{\ds} \BLUE{\C^{33}}) {\ds} (\GREEN{\C^{30}}{\ds} \VIOLET{\C^{30}}){\ds} (\RED{\C^{30}}{\ds} \GRAY{\C^{30}}){\ds} \MAG{\C^{30}}{\ds} \BLUE{\C^{21}} {\ds} \GREEN{\C^{13}}{\ds} \VIOLET{\C^{6}}$ &$\MAG{\C^{30}}$&$\MAG{\C^{30}}$&$\MAG{\C^{30}}$
\\[1mm]
9&  $(\RED{\C^{58}}{\ds} \RED{\C^{58}}) {\ds} (\GREEN{\C^{56}}{\ds} \GREEN{\C^{56}}){\ds} (\VIOLET{\C^{56}}{\ds} \VIOLET{\C^{56}}){\ds} (\RED{\C^{56}}{\ds} \RED{\C^{56}}){\ds} \GREEN{\C^{46}}{\ds} \VIOLET{\C^{14}}$ &$\GREEN{\C^{46}}$&$(\VIOLET{\C^{56}}{\ds} \VIOLET{\C^{56}})$&$\GREEN{\C^{46}}$
\\[1mm]
10 &  ($\RED{\C^{105}}{\ds} \RED{\C^{105}} ){\ds} (\GREEN{\C^{105}}{\ds} \GREEN{\C^{105}}){\ds} (\GRAY{\C^{99}}{\ds} \GRAY{\C^{99}}){\ds} (\VIOLET{\C^{99}}{\ds} \VIOLET{\C^{99}}){\ds} \MAG{\C^{78}} {\ds} \RED{\C^{58}}{\ds}\VIOLET{\C^{42}} {\ds} \MAG{\C^{30}}$ &$\MAG{\C^{30}}$&$\MAG{\C^{30}}$&$\MAG{\C^{30}}$
\\[1.5mm]
\hline\hline
\end{tabular}
\caption{Reachability of the Heisenberg and Ising target state for the Rydberg-atom platform of Sec.~\ref{sec:VQE}. The invariant subspaces are determined together with the support of the initial state and the Heisenberg and Ising target ground state(s) for three to ten qubits. 
We use the color scheme from Table~\ref{table:symmetry}.
For the Heisenberg case, the initial and the target states are in different isotypic components, except for four, eight, and ten qubits. This implies that the Heisenberg target state can never be reached for three, five, six, seven, and nine qubits using VQE starting from the chosen initial state. For example, for five qubits, one observes the three invariant subspaces $\RED{\C^8}{\ds} \BLUE{\C^6} {\ds} \GREEN{\C^6}{\ds} \VIOLET{\C^6}{\ds} \GRAY{\C^6}$. The initial state has only support in the first one $\RED{\C^8}$  and the target state has only support in the second and third one
$\BLUE{\C^6}$ and $\GREEN{\C^6}$.
\label{table_app:symmetry}}
\end{table*}

We provide further details for the Lie-algebra computations 
following the description in Sec.~\ref{lie_algebra}. All computations
are implemented using the 
computer algebra system \textsc{Magma} \cite{magma1997}
which enables computations with exact fields such as the rational field
or number fields (including cyclotomic fields) \cite{cohen1993}.
Recall that number fields are field extensions of finite degree
of the rational field.
Computations with exact fields are complemented with
numerical computations based on inexact fields
such as (approximated) real or complex fields.
Often computations over the (not approximated) real field
can be substituted with exact computations
over the rational field or a suitable number field.
The employed computational techniques
in this work (and particularly in the Appendices~\ref{app:lie_algebra}
and \ref{app:symmetry_analysis})
build on and partially
extend computational tools applied in
\cite{zeier_symmetry_2011,zimboras_dynamic_2014,zeier2015squares,zimboras_symmetry_2015,schulte2017,schulte2018}
and, most recently, in \cite{kazi_analyzing_2024}.

Starting from a set of (hermitian) generators $\gens$ (such as the resource Hamiltonians $\gens_R=\{H_d, H_\Delta, H_\Omega\}$),
we can compute the Lie-algebra closure $\g:=\lie{\gens}$ by repeatedly computing commutators 
(starting from the elements $iH_j$ for $H_j$ in $\gens$)
and determining the (linear-algebra) rank of a set of Lie-algebra elements.
For the Lie-algebra closure, $\g$ is considered as a Lie subalgebra of $\su(2^n)$ [or $\uu(2^n)$]
which is specified by its structure constants \cite{deGraaf2000} over a Pauli-string basis,
where Pauli strings are given by $n$-fold tensor products of Pauli operators $\sigma_j$ with $j\in \{x,y,z\}$
and the $2\times 2$ identity matrix $\unity_2$.
If the generators can be determined as linear combinations 
over the rational field (or a suitable number field contained in the real field),
the generated Lie algebra can be exactly calculated as
all rank computations can be exactly performed.

Based on a basis for $\g$ (as detailed above), we can determine
its reductive decomposition $\g= \g_1\rd \cdots \rd \g_k$ into simple or abelian components $\g_j$
using \textsc{Magma} \cite{deGraaf2000,magma1997} by first computing the components $\g_j$ and
then identifying the simple components as one of the classical simple Lie algebras $\su(p)$ with $p\geq 2$,
 $\so(2p{+}1)$, $\usp(p)$, and $\so(2p)$ for suitable integers $p\geq 1$
or one of the five exceptional ones \cite{hall2015,deGraaf2000}. But we are referring only to the corresponding compact real forms
as all considered Lie algebras are compact \cite{hall2015}, i.e., they are contained in $\uu(2^n)$.

Table~\ref{table:lie_algebra} details the reductive decompositions for the 
Lie algebra $\g_R$ generated by the resource Hamiltonians $\gens_R=\{H_d, H_\Delta, H_\Omega\}$
for three to seven qubits. Similarly, Table~\ref{table:lie_algebra} also specifies
the Lie algebras $\h_H$ and $\h_I$ obtained when the generating set $\gens_R$ is extended by the
Heisenberg and the Ising Hamiltonians $\target_H$ and $\target_I$ [see Eqs.~\eqref{eqn:heisenberg} and \eqref{eqn:ising}]. 
We observe in Table~\ref{table:lie_algebra} that the Lie algebras $\h_H$ and $\h_I$ are isomorphic. In particular,
the Lie algebras $\g_R$, $\h_H$, $\h_I$ are isomorphic for three, five, and seven qubits. But, for four and six qubits, 
$\g_R$ differs from $\h_H$ and $\h_I$  by
an element of the Lie center in $\h_H$ and $\h_I$, where
the dimension of the Lie centers of $\g_R$ and respectively $\h_H$ and $\h_I$  are one and  two.

\section{Symmetry computations\label{app:symmetry_analysis}}

In this appendix, we 
provide further details relevant to the symmetry analysis discussed
in Sec.~\ref{symmetry_analysis} while highlighting computational aspects
as in Appendix~\ref{app:lie_algebra}.
In a first step, we compute the commutant 
of the resource Hamiltonians $\gens_{R}=\{H_d, H_\Delta, H_\Omega\}$
(see Sec.~\ref{symmetry_analysis}) and the commutant will now be denoted
by $\mathcal{C}$.

Based on the commutant $\mathcal{C}$, we can determine 
the Lie-algebra center $\Z(\g_R)=\mathcal{C} {\cap} \g_R$
of $\g_R$ without computing $\g_R$. Consider a skew-hermitian basis
of the commutant with basis elements $b_k$ and $k \in \{1,\ldots,\dim(\mathcal{C})\}$.
We project the resource Hamiltonians $H_j$ onto this basis using the
formula $i \mathrm{Tr}(H_j b_k)/\mathrm{Tr}(b_k b_k)$,
where $\mathrm{Tr}(M)$ denotes the trace
of a matrix $M$. A basis of the projected operators, which are expanded as a linear combination of
the elements $b_k$, then forms a basis of the Lie center $\Z(\g_R)$.
If one is only interested in the dimension of $\Z(\g_R)$, one can compute the rank
of the matrix $R$ with entries $R_{ij}=\text{Trace}(C_i^\dagger H_j)$, where
the $C_j$ describe any basis of $\mathcal{C}$ and the matrix $R$ has $\abs{\gens_R}$ columns
and $\dim(\mathcal{C})$ rows (see \cite{zimboras_symmetry_2015}).

We can compute the isotypic and irreducible projectors as outlined in Sec.~\ref{symmetry_analysis}.
We detail now one particular optimization for determining the irreducible projectors
by first reducing the block size of the resource Hamiltonians
with respect to a considered isotypic projector $P_j$
[see Fig.~\ref{fig:overview}(a)]. Following some idea in \cite{kazi_analyzing_2024},
we can construct rectangular matrices $Q_j$ and $Q^\dagger_j$, 
where the columns of $Q^\dagger_j$ are given by an orthonormal eigenbasis of $P_j$.
We use the Gram-Schmidt process 
for the orthogonalization and then normalize each eigenvector
$\nu_j$ by $\nu'_j=\nu_j/\vert \nu_j \vert$, where $\vert \nu_j \vert = \sqrt{\langle \nu_j \vert \nu_j \rangle}$
and field extensions are required for the square roots. Hence, we can write the matrix $Q_j$ as $Q_j = D_{\mathrm{sqrt}}Q'_j $, where $Q'_i$ contains the orthogonal eigenbasis of $P_j$ and $D_{\mathrm{sqrt}}$ is a diagonal matrix containing all the normalization factors $\vert \nu_j \vert$.
Similarly, we can write $Q^\dagger_j=Q'^\dagger_j D_{\mathrm{sqrt}}$. We then apply the formula
$\tilde{H}_v=Q_jH_vQ^\dagger_j$ for $H_v \in \gens_R$. Consequently, the meataxe algorithm
\cite{lux2010} mentioned in Sec.~\ref{symmetry_analysis} can be applied to smaller matrices.

This then enables us to compute the projections of state vectors $s$ via $s P_j$ for an isoyptic subspace or $s Q^\dagger_j \tilde{P}_j$
for an irreducible subspace. For Hamiltonians $W$ or Lie algebra elements $iW$, we similarly have the formulas
$P_j W P_j$ and $\tilde{P}_jQ_j W Q^\dagger_j \tilde{P}_j$. Thus we can determine the non-zero support.
In summary, Table~\ref{table_app:symmetry} details the invariant subspaces and the non-zero 
support of the initial and the target state(s) for three to ten qubits.

\section{Lie centers when adding the target Hamiltonian to the resource Hamiltonians \label{app:lie-center}}

\begin{table}[t]
\centering
\begin{tabular}{l@{\hspace{7mm}} l @{\hspace{6mm}} c @{\hspace{7mm}} c@{\hspace{6mm}} r @{\hspace{4mm}} r}
\hline\hline
\\[-2mm]
& \multicolumn{2}{@{}l}{Lie center} & \multicolumn{3}{r}{Impact from the }\\
Number &  \multicolumn{2}{@{}l}{dimension for}  & \multicolumn{3}{r}{extended center for} \\
of qubits&
$\g_R$ & $\h_H$ and  $\h_I$ && $\h_H$ & $\h_I$
\\[1mm]
\hline
\\[-2mm]
3 &  1&1&&N/A&N/A
\\[1mm]
4 &  1&2&&None&None
\\[1mm]
5 &  1&1&&N/A&N/A
\\[1mm]
6 &  1&2&&Possible&None
\\[1mm]
7 &  1&1&&N/A&N/A
\\[1mm]
8 &  1&2&&None&None
\\[1mm]
9 &  1&2&&Possible&None
\\[1mm]
10 &  1&2&&None&None
\\[1.5mm]
\hline\hline
\end{tabular}
\caption{The impact of additional Lie center element(s) on the reachability of the target ground state(s)
when extending the resource Hamiltonians with the respective Heisenberg or Ising target Hamiltonians
$\target_H$ or $\target_I$. This then results in the extended Lie algebras $\h_H$ or $\h_I$.
A potential impact on the VQE is shown by ``Possible'' and no effect by ``None"; ``N/A'' indicates
the non-existence of additional center elements. Refer to the text of Appendix~\ref{app:lie-center}
for further details.
\label{table:add}}
\end{table}

In this Appendix, we discuss whether  adding
the target Hamiltonian to the resource Hamiltonians
would affect the result of the reachability analysis.
The Lie algebras $\g_R$, $\h_H$, and $\h_I$ without and with the target Hamiltonians and their reductive decompositions 
have been computed for up to seven qubits. The results are
detailed in Table~\ref{table:lie_algebra} of Appendix~\ref{app:lie_algebra} and they suggest
that the extended Lie algebras $\h_H$ and $\h_I$ differ from the resource Lie algebra $\g_R$
by at most one additional center element. So, our analysis beyond seven qubits is predicated by
general correctness of this observation.
The Lie-algebra center dimension is calculated without and with the Heisenberg and the Ising target Hamiltonian. 
We observe in certain cases an extended Lie center which is reflected in a larger 
Lie-center dimension as recorded in column three of Table~\ref{table:lie_algebra}.
These additional center elements could potentially lead to an extended 
reachability.

We can effectively compute the Lie centers via the commutant
beyond seven qubits and analyze the support of the corresponding center elements.
Table~\ref{table:add} summarizes the results by detailing the Lie-center dimensions 
as well as the potential impact on the isotypic subspaces with non-zero support
for the initial and the target states.
An additional center element is observed for four, six, eight, nine, and ten qubits.
They have no impact on the initial and the target state in the invariant
subspace $\C^6$, $\C^{30}$, and $\C^{30}$ for four, eight, and ten qubits respectively.
This means that time evolutions enabled by the additional center element do not affect
the invariant subspace containing the initial and the target state.

However, the initial and the target states are located in the different invariant subspaces $\C^{13}$ and $\C^7$ for six qubits
as shown in Table~\ref{table:symmetry}. The additional center element does not act on the initial state, but it does act
on the target state. These observations do not change our assessment of the reachability of the Heisenberg target ground state.
A similar argument applies to the case of nine qubits. For the Ising model, the additional center element has zero support
in the isotypic subspaces that have nonzero support for the initial state and the target state.
We conclude from Table~\ref{table:add} that the
additional center elements cannot affect the VQE for the Ising model with all considered qubits.
Thus adding the target Hamiltonian to the resource Hamiltonians
appears to not help to reach the target ground state for up to ten qubits.


%

\end{document}